# Magnetic Structure of CaBaCo$_4$O$_7$:

# Lifting of Geometrical Frustration towards Ferrimagnetism

V. Caignaert[1,*], V. Pralong[1], V. Hardy[1], C. Ritter[2] and B. Raveau[1]

*(1)CRISMAT, UMR 6508, CNRS-ENSICAEN*
*6 Bd Marechal Juin, 14050 Caen, France*
*(2) Institut Laue Langevin, 6 rue Jules Horowitz, 38042 Grenoble, France*

**Abstract**

CaBaCo$_4$O$_7$ represents a new class of ferrimagnets whose structure is built up of CoO$_4$ tetrahedra only, similarly to other members LnBaCo$_4$O$_7$ of the "114" series, forming an alternate stacking of kagomé and triangular layers. Neutron powder diffraction reveals, that this compound exhibits the largest distortion within the "114" series, characterized by a strong buckling of the kagomé layers. Differently from all other members it shows charge ordering, with Co$^{2+}$ sitting on two sites (Co2, Co3) and "mixed valent" cobalt "Co$^{3+}$/Co$^{2+}$L" sitting on two other sites (Co1, Co4). The unique ferrimagnetic structure of this cobaltite at 4 K can be described as the assemblage of ferrimagnetic triple chains (Co1 Co2 Co3) running perpendicular to the kagomé layers, ferromagnetically coupled within the layers, and antiferromagnetically coupled with a fourth cobalt species Co4. The lifting of the geometrical frustration towards ferrimagnetism, which appears in spite of the triangular topology of the cobalt lattice, is explained by the very large structural distortion, charge ordering phenomena and large cobalt valence compared to other LnBaCo$_4$O$_7$ oxides.

* Corresponding author: Dr. Vincent Caignaert

e-mail: vincent.caignaert@ensicaen.fr

Fax: +33 2 31 95 16 00

Tel:  +33 2 31 45 26 32

The competition between geometrical frustration and magnetic ordering in transition metal oxides leading to exotic magnetic properties and transitions has been the subject of many investigations these last few decades [1] that are still to date not completely understood. In this respect, oxides whose metallic sublattice exhibits vertex-sharing triangular or tetrahedral topology with antiferromagnetic (AFM) interactions are of great interest. This is for example the case in $A_2B_2O_7$ pyrochlores [2-3], whose metal framework consists of tetrahedral $[A_4]_\infty$ or $[B_4]_\infty$ sublattices. The spinels $AB_2O_4$ may be also geometrically frustrated magnets when the A cation is diamagnetic [4]. A second important example deals with the "kagomé" oxides, where the magnetic frustration originates from a 2D triangular topology of their magnetic network. This class of compounds is nicely illustrated by the Jarosite family [5], which display either a spin-glass behavior or long-range magnetic order with propagation vectors $k=0$ and $k=(1/3, 1/3, 0)$.

The discovery, some years ago of the cobaltites $LnBaCo_4O_7$ [6-9] and very recently of the ferrites $CaBaFe_4O_7$ [10] and $YBaFe_4O_7$ [11, 12] opened the route to the study of new geometrically frustrated magnets. Structurally, these compounds are closely related to the spinels as far as they contain similar triangular sublattices of cobalt or iron atoms with the kagomé topology, but they differ fundamentally from the latter by the fact that their Co-O or Fe-O framework consists exclusively of $CoO_4$ or $FeO_4$ tetrahedra, forming two sorts of layers, called triangular and kagomé respectively. As a consequence, the magnetic properties of these oxides are very different from those of other "kagomé" oxides and show very complex transitions. Such a behavior is exemplified by the cobaltite $YBaCo_4O_7$ for which a spin glass transition was first reported around $T_f \sim 66$ K [7], whereas long range magnetic order was evidenced below $T_N = 110$K [13], and a magnetic transition with short range correlations was revealed above $T_N$ in a single crystal of this phase [15]. From the structural view point, this phase also exhibits a structural transition from orthorhombic to hexagonal above room temperature at $T_S = 313$ K [13], the orthorhombic low temperature form corresponding to a distortion of the high temperature hexagonal form. Quite remarkably, all the magnetic transitions appear below $T_S$ and the recent studies of $YBaCo_4O_7$ by single crystal neutron scattering by Manuel et al. [15] demonstrate a very important feature: At 130 K, i.e. above $T_N$, this phase exhibits a quasi 1D magnetic order along $\vec{c}$, whereas in the kagomé layers (a-b plans) it displays a strong degeneracy, so that it can be described as a new class of 2D frustrated oxide.

Recently, we synthesized the cobaltite $CaBaCo_4O_7$ [16], which differently from all other cobaltites of this series, including $Y_{0.5}Ca_{0.5}BaCo_4O_7$ [17], exhibits ferrimagnetic properties below $T_C = 70$ K. Bearing in mind that the orthorhombic distortion of this phase is much larger than any observed for other cobaltites, we have explored its nuclear and magnetic structure. We report herein on the crucial role of the structural distortion, of the cobalt valence and of charge ordering in the appearance of ferrimagnetic ordering in this phase. We show that the AFM out of plane coupling in the "Co1 Co2 Co3" triple chains running along $\vec{c}$ is a key parameter for weakening the in-plane magnetic frustration and consequently for inducing the 3 D magnetic ordering.

The sample was synthesized from a stoichiometric mixture of $CaCO_3$, $BaCO_3$ and $Co_3O_4$ first heated at 900°C in air for 12 h for decarbonation. The mixture was then heated in air at 1100°C for 12 h, and quenched down to room temperature. Specific heat measurements were carried out by means of a commercial Physical Properties Measurements System (PPMS, Quantum Design) using a relaxation method with a 2τ fitting procedure. Neutron powder diffraction (NPD) versus temperature was performed on the D2b diffractometer (ILL, Grenoble) at room temperature (λ = 1.59Å) and from 4 to 150 K (λ = 2.428 Å).

The NPD patterns registered at various temperatures reflect an excellent crystallization of the compound, as illustrated for the pattern of $CaBaCo_4O_7$ registered at 4 K (Fig.1). An impurity, identified as CoO, was taken into account, due to the presence of one magnetic reflection (2θ=28.2°, d=4.92Å) close to those of $CaBaCo_4O_7$. The amount of CoO was estimated to 2% in weight and is probably related to $Co_3O_4$ in excess during the synthesis. Whatever the temperature, the Rietveld refinements of $CaBaCo_4O_7$ evidence the orthorhombic $Pbn2_1$ symmetry, in the whole temperature range. This is in contrast to most of the other $LnBaCo_4O_7$ cobaltites [8-9] which generally show a structural transition from orthorhombic (O) to hexagonal (H) symmetry as the temperature increases, corresponding to the relations: $a_O \sim a_H \sim 6.3$ Å, $b_O \sim a_H\sqrt{3} \sim 11$ Å and $c_O \sim c_H \sim 10.2$ Å. The evolution of the cell parameters versus temperature (Fig.2a) shows that the $\vec{a}$ and $\vec{c}$ parameters decrease with T, whereas the $\vec{b}$ parameter increases as T decreases, leading to an overall decrease of the cell volume (Fig.2b). The orthorhombic symmetry results from a distortion of the hexagonal cell, which can be quantified as $D = (b/\sqrt{3} - a)/a$, corresponding to an expansion of the hexagonal cell along one $[110]_H$ direction and a contraction along the $[1\bar{1}0]_H$ perpendicular direction. The amplitude of this distortion D in $CaBaCo_4O_7$ is the largest ever observed in the "114" series of cobaltites. It increases linearly as T decreases from D=1.05% at room temperature to D=1.8% at 50 K, and remains constant below this temperature down to 4 K (Fig.2b). These values can be compared to the D values previously observed for the $LnBaCo_4O_7$ compounds at low temperature. For the latter, they are always much smaller, reaching a maximum value of 0.5% at 10 K for $YbBaFe_4O_7$, i.e. still much smaller than the D value observed for $CaBaCo_4O_7$ at room temperature. This high value of the distortion is quite remarkable, since it induces a lifting of the geometrical frustration, which is initially present in the cobalt network of the LnBaCo$_4$O$_7$ series [6-9], due to its triangular topology. It explains the absence of magnetic frustration at low temperature, and the appearance of a ferrimagnetic state in CaBaCo$_4$O$_7$.

The atomic coordinates of this oxide, determined at 4 K (Table 1) and at room temperature by NPD are in agreement with the results obtained from X-Ray powder diffraction refinements [16], but show a higher accuracy of the oxygen positions. The crystal structure of CaBaCo$_4$O$_7$ consists of a 1:1 stacking of kagomé (K) and triangular (T) layers of CoO$_4$ tetrahedra along $\vec{c}$. The projections of the structure along $\vec{a}$ (Fig.3a) and $\vec{b}$ (Fig.3b) show that the triangular layers (T) formed by the Co1 tetrahedra are perfectly flat; in contrast, the kagomé layers (K) are strongly corrugated, the apical oxygen of the Co2, Co3 and Co4 tetrahedra being located in a plane, whereas the oxygen atoms of the basal planes of these tetrahedra form a waving layer.

The interatomic distances (Table 2) show that the four symmetry-independent cobalt tetrahedra can be classified into two groups: the Co2 and the Co3 tetrahedra belonging to the K layers which exhibit larger Co-O distances, i.e. average distances ranging from 1.94-1.97 Å (RT) to 1.95 Å (4 K), and the Co1 and Co4 tetrahedra belonging to the T and K layers respectively, whose Co-O distances are significantly smaller i.e. average distances ranging from 1.87-1.88 Å (RT) to 1.85-1.89 Å (4 K). Such a distribution of the Co-O bond lengths between these four independent crystallographic sites suggests that, whatever the temperature, CaBaCo$_4$O$_7$ exhibits charge ordering in agreement with the stoichiometric formula CaBaCo$_2^{2+}$Co$_2^{3+}$O$_7$. Nevertheless, though the bond lengths observed for Co2 and Co3 are in agreement with those observed for Co$^{2+}$ in tetrahedral coordination, the Co-O bond lengths observed for Co1 and Co4 are significantly larger than those expected for Co$^{3+}$ in tetrahedral coordination, which were found typically close to 1.79 Å in oxides. The bond valence sum calculations (BVS), performed according to Alternatt and Brown [19], support this viewpoint (Table 3). One indeed observes that Co2 and Co3 exhibit a charge comprised between +1.9 and +2.1, whatever the temperature and can be considered as divalent, whereas a charge ranging from +2.4 to +2.6 is observed for Co1 and Co4, suggesting that these cations should be rather mixed valent, with an average value of +2.5. Such a mixed valence, significantly smaller than +3, is in contradiction with the stoichiometric formula. Indeed, from structure refinements neither oxygen deficiency leading to the formula CaBaCo$_4$O$_{6.5}$, nor presence of protons according to the formula CaBaCo$_4$O$_6$OH were detected. Bearing in mind that the tetrahedral coordination of Co$^{3+}$ is very rare, due to the crystal field effect which favors the octahedral coordination, the possibility of partial charge transfer from cobalt to one oxygen may be considered. This would lead to a ligand hole configuration (L̲) similar to that observed for apical oxygen in layered cuprates [20], corresponding to a hybridization of the bonds of the Co1 and Co4 tetrahedra according to the scheme Co$^{3+}$-O = (Co3d$^6$) ⇔ Co$^{2+}$=O-(Co3d$^7$L̲). The significantly larger value of the apical Co-O7 bond length, of the Co1 and Co4 tetrahedra, ranging from 1.91-1.95 Å at RT to 1.97 – 1.89 Å at 4 K, suggests that the hole appears on the O7 atom shared between one Co4 and one Co1 tetrahedron. A study of this compound by X-ray absorption spectroscopy would be necessary to elucidate this point.

The geometry of the $CoO_4$ tetrahedra is significantly modified by temperature. The size of the Co1 and Co2 tetrahedra increases as T decreases, whereas the size of the Co3 and Co4 tetrahedra decreases with temperature, as shown from the average <Co-O> bond lengths (Table2). The evolution of the distortion is as well very complex. The distortion of the Co1 and Co3 tetrahedra increases significantly as T decreases, whereas the distortion of the Co2 tetrahedra decreases slightly with T, and no significant variation of the distortion with temperature is observed for the Co4 tetrahedra. Clearly, the structural evolution vs T does not influence the charge ordering of $CaBaCo_4O_7$: it can be described as 1:1 charge ordering between Co2/Co3 sites occupied by $Co^{2+}$ and Co1/Co4 sites which are mixed or hybridized $Co^{3+}$ ($3d^6$) / $Co^{2+}$ ($3d^7\underline{L}$).

The Ca-O distances of the $CaO_6$ octahedra, ranging from 2.25 to 2.40 Å at 4 K and from 2.27 to 2.35 Å at RT are in agreement with those usually observed. The coordination of barium changes from 5+7 at RT to 6+6 at 4 K. The 5 or 6 nearest oxygen neighbors are located at usual distances from barium, ranging from 2.80 to 2.90 Å, whereas the other seven or six oxygen atoms sit much further, at distances ranging from 3.36 to 3.58 Å. In spite of this reasonable coordination of barium, the computed valence of barium calculated from BVS is much smaller than expected i.e. +1.4 instead of +2, suggesting that $Ba^{2+}$ is strongly underbonded. A similar behavior has been observed previously for $YbBaCo_4O_7$ [18] and was attributed to a structural instability, in agreement with the transition from hexagonal to orthorhombic symmetry at decreasing temperature. Such a hypothesis does not hold here, since $CaBaCo_4O_7$ does not exhibit any structural transition in the whole temperature range and moreover the so-obtained barium valence of 1.40 for $CaBaCo_4O_7$ at 4K, which exhibits the largest distortion, is still very close to that obtained for the hexagonal form of $YbBaCo_4O_7$ (1.33). The origin of this abnormally weak valence of barium remains unexplained.

The NPD pattern registered at 4 K shows clearly the presence of magnetic Bragg peaks (Fig.1) with a propagation vector k=0, whose intensity decreases as T increases. The latter disappear above $T_C$=70 K, in agreement with the specific heat measurement (inset Fig.1) and with the ferrimagnetic to paramagnetic transition previously observed for this oxide [16]. To generate all the spin configurations compatible with the crystal symmetry we carried out a group theory analysis using the program BasiReps. There are four irreducible representations (IR) associated with the *Pbn2₁* space group and *k*=(0, 0, 0). Among the latter, only three IRs allow a ferrimagnetic alignment of Co sublattices. The basis vectors of these representations are listed in Table 4. The results of simulated annealing runs, to solve the magnetic structure, lead to a unique solution in the $\Gamma_4$ representation. The final refinement shows that one of the 3 coefficients ($\Psi_{12}$ in Table 4), representing the 3 basis vectors, refines to zero leading to a magnetic structure where all the spins lie in the (*ab*) plane. This solution of the magnetic structure at 4 K confirms its ferrimagnetic nature, characterized by a simultaneous ordering of Co spins in both triangular and kagomé layers. Thus, the magnetic structure at 4 K (Fig.4) can be described in the *Pbn2$_1$* space group, with all Co spins lying in the (a,b) plane, i.e. parallel to the kagomé and triangular layers. The ferrimagnetic structure of this cobaltite is quite unique with respect to all other members of the LnBaCo$_4$O$_7$ series. In the kagomé layers, the Co2 and Co3 spins form zigzag chains running along $\vec{b}$ (Fig.4a and 7b). In those chains, one Co2 alternates with one Co3 species, and two successive Co2 (or Co3) spins are oriented at ~60°, while "Co2Co3" ferromagnetic couples are formed. As a result, the transverse magnetic components of Co2 (and Co3) along $\vec{a}$ are antiferromagnetically coupled and cancel each other, whereas the longitudinal magnetic components of Co2 and Co3 along $\vec{b}$ are all oriented in the same direction. Thus, Co2 and Co3 form ferromagnetic zigzag chains along $\vec{b}$, which are themselves all ferromagnetically coupled in the entire kagomé layer. Between those chains, the Co4 species have their spins oriented antiparallel to the FM resultant of "Co2Co3" chains. Consequently, a ferrimagnetic order of the Co2, Co3 and Co4 spins takes place in the kagomé layers, whose facile magnetization is directed along $\vec{b}$. The Co1 spins of the triangular layers (Fig.4a-b) are practically antiparallel to Co2 and Co3 spins (Fig.4a and 7b), two successive Co1 spins, being oriented at 60°, similarly to "Co2 Co3" chains. Thus, their transverse components are antiparallel and cancel each other and their longitudinal components are parallel, forming ferromagnetic chains Co1 running along $\vec{b}$. The latter chains are antiferromagnetically coupled with the "Co2 Co3" chains. In summary, the 3 D ferrimagnetic ordering of cobalt spins in CaBaCo$_4$O$_7$, consists of "Co2 Co3" zigzag FM chains running along $\vec{b}$, antiferromagnetically coupled with Co4 and Co1 spins along $\vec{a}$ and $\vec{b}$ respectively, the resultant magnetic moments of cobalt spins being parallel to the facile magnetization axis $\vec{b}$. Bearing in mind the description of the cationic framework of CaBaCo$_4$O$_7$, previously proposed by Chapon et al [13], which consists of chains of corner-sharing "Co5" bipyramids running along $\vec{c}$ (Fig.4b) interconnected through "Co3" triangles in the (a,b) plane (Fig.4a), another description of this ferrimagnetic structure can be proposed. It consists of the assemblage of ferrimagnetic chains running along $\vec{c}$, and corresponding to the AFM coupling of Co1, with Co2 and Co3. In this description, each triple "Co1 Co2 Co3" chain is ferromagnetically coupled to two other identical chains, and antiferromagnetically coupled to three Co4 species in the (a, b) plane.

The evolution of the spontaneous magnetization versus temperature (Fig.5) corroborates the previous magnetic measurements [16]. One indeed observes a resultant magnetic moment of ~ $1\mu_B$ per f.u. at low temperature, close to the value observed from M(T) data (~ $0.6\mu_B$ per f.u.), which decreases abruptly at $T_C$ down to ~$0.2\mu_B$ .

The evolution of the magnetic moments of the four cobalt sites, versus temperature (Fig.6), shows also an abrupt decrease around $T_C$ as expected. More importantly, the values of these moments at low temperature support strongly the existence of charge ordering. One indeed observes significantly larger values of the magnetic moments of Co1 and Co4 (2.5 - 3.0 $\mu_B$ per Co) in agreement with their valences deduced from BVS's, i.e. $Co^{2+}\underline{L}/Co^{3+}$ (theoretical moment ~ $4\mu_B$). These values of the magnetic moments, smaller than the theoretical ones, suggest that a degree of frustration remains, inducing a slight disordering of the cobalt spins.

This study shows that $CaBaCo_4O_7$ exhibits, compared to all other "114" $LnBaCo_4O_7$ cobaltites, a very unique ferrimagnetic structure, in spite of its close structural relationships with these oxides. The origin of this complex magnetic structure is, most likely, reminiscent of the competition between geometric frustration that appears in the kagomé layers due to their triangular topology and the antiferromagnetic interactions that take place along $\vec{c}$ of corner-shared tetrahedra (Co1, Co2 and Co3). The crucial role of these triple AFM cobalt chains in the establishment of the 3 D magnetic ordering in these compounds is clearly shown by comparing the magnetic structure of $YBaCo_4O_7$ previously observed at 80 K by Chapon et al [15] (Fig.7a) with the structure of $CaBaCo_4O_7$ at 4 K (Fig.7b). Both magnetic structures consist of very similar triple AFM chains "Co1 Co2 Co3" running along $\vec{c}$ and ferromagnetically coupled along $\vec{a}$, i.e. forming ferrimagnetic layers parallel to (010). The two structures differ by the coupling of those layers along $\vec{b}$, which is antiferromagnetic for $YBaCo_4O_7$ (Fig.7a) and ferromagnetic for $CaBaCo_4O_7$ (Fig.7b). The spin orientation of Co4 is also different in the two structures, the moment of Co4 being oriented at 90° from other spins in $YBaCo_4O_7$ and at ~ 30° in $CaBaCo_4O_7$. The Co4 spins are antiferromagnetically coupled in $YBaCo_4O_7$ (Fig.7a) whereas they are ferromagnetically coupled along $\vec{b}$ in $CaBaCo_4O_7$ (Fig.7b), so that the magnetic structure of the latter phase remains ferrimagnetic with $\vec{b}$ as facile magnetization axis. In both structures, the lifting of the magnetic frustration is explained by the fact that the in-plane exchange interaction $J_1$, which favors the 120° geometry in the hexagonal structure, is reduced by the orthorhombic distortion as T decreases. This orthorhombic distortion is much larger for $CaBaCo_4O_7$, as shown from the in-plane Co-Co distances between Co2, Co3 and Co4 ranging from 3.07 to 3.33 Å at RT and from 3.00 to 3.31 Å at 4 K, so that the Co triangles are no more equilateral. The 3 D magnetic ordering is also supported by the out of plane $J_2$ interaction, which tends to favor the existence of AFM triple chains "Co1 Co2 Co3" in both structures. The fact that the cobalt valence is larger in $CaBaCo_4O_7$, but especially that $Co^{3+}$/or $Co^{2+}\underline{L}$ are preferentially located on the Co1 and Co4 sites may change significantly $J_2$. Such a charge ordering may be at the origin of the FM coupling of the chains in $CaBaCo_4O_7$, in contrast to the AFM coupling observed for $YBaCo_4O_7$. Thus, the combination of these different factors – structural distortion, mixed valence of cobalt and charge ordering – appear as key factors for controlling the magnetic ordering in the "114" cobaltite series.

    Finally, it is worth pointing out that the ferrimagnetism observed for $CaBaCo_4O_7$, is significantly different from what has been observed for $CaBaFe_4O_7$ [10] where $T_C$ and magnetic moments are much higher. The hexagonal symmetry of the latter suggests that the iron spins in this phase might be lying out of plane.

**Table Captions:**

**Table1:** Atomic and magnetic parameters for $CaBaCo_4O_7$ at 4K. All sites are fully occupied. Space group: *Pbn2₁*. Cell parameters: a= 6.2613(1) Å, b=11.0399(2) Å, c=10.1642(2) Å.

| Atom | x | y | z | B |
|---|---|---|---|---|
| Ca | 0.0066(21) | 0.6729(13) | 0.8708(13) | 0.06(13) |
| Ba | 0.0043(14) | 0.6651(10) | 0.5 | 0.06(13) |
| Co1 | 0.0155(33) | -0.0003(24) | 0.9332(25) | 0.09(13) |
| Co2 | -0.0067(26) | 0.1688(20) | 0.6902(28) | 0.09(13) |
| Co3 | 0.2527(33) | 0.0955(16) | 0.1904(21) | 0.09(13) |
| Co4 | 0.2673(25) | 0.9209(15) | 0.6841(21) | 0.09(13) |
| O1 | -0.0155(27) | 0.0044(11) | 0.2460(16) | 0.64(5) |
| O2 | -0.0045(20) | 0.4914(11) | 0.2285(16) | 0.64(5) |
| O3 | 0.7817(21) | 0.2613(10) | 0.7815(19) | 0.64(5) |
| O4 | 0.7319(18) | 0.7418(12) | 0.2141(17) | 0.64(5) |
| O5 | -0.0549(12) | 0.1522(8) | 0.4997(16) | 0.64(5) |
| O6 | 0.2035(13) | 0.1097(6) | 0.0062(19) | 0.64(5) |
| O7 | 0.2645(15) | 0.9471(6) | 0.5002(22) | 0.64(5) |

| Atom | $M_x$ | $M_y$ | M |
|---|---|---|---|
| Co1 | -1.78(16) | -2.20(16) | 2.83(16) |
| Co2 | 1.01(14) | 1.77(20) | 2.04(19) |
| Co3 | 0.72(22) | 1.92(11) | 2.05(15) |
| Co4 | -0.05(20) | -2.44(12) | 2.44(15) |

**Table 2:** Interatomic distances at 4K for $CaBaCo_4O_7$.

| Bond | Distance (Å) | Bond | Distance (Å) |
|---|---|---|---|
| Ca-O2 | 2.32(4) | Co1-O1 | 1.90(6) |
| O3 | 2.24(4) | O5 | 1.82(5) |
| O4 | 2.32(4) | O6 | 1.85(5) |
| O5 | 2.35(4) | O7 | 1.98(5) |
| O6 | 2.38(4) | Co2-O1 | 2.00(5) |
| O7 | 2.41(4) | O3 | 1.91(5) |
| Ba-O2 | 3.36(3) | O4 | 2.00(4) |
| O2 | 2.89(3) | O5 | 1.97(6) |
| O3 | 3.54(4) | Co3-O1 | 2.04(5) |
| O3 | 2.72(4) | O2 | 1.99(5) |
| O4 | 3.48(3) | O3 | 1.83(5) |
| O4 | 2.79(3) | O6 | 1.90(6) |
| O5 | 2.82(2) | Co4-O1 | 1.89(5) |
| O5 | 3.45(2) | O2 | 1.78(4) |
| O6 | 3.57(3) | O4 | 1.84(4) |
| O6 | 2.81(3) | O7 | 1.89(6) |
| O7 | 3.51(3) | | |
| O7 | 2.81(3) | | |

**Table 3**: Bond valence sum calculations (BVS) at 4K.

| Atom | BVS |
|---|---|
| Ca | 2.23(5) |
| Ba | 1.48(3) |
| Co1 | 2.39(8) |
| Co2 | 1.89(7) |
| Co3 | 2.08(7) |
| Co4 | 2.63(9) |
| O1 | 1.97(7) |
| O2 | 1.82(6) |
| O3 | 2.01(7) |
| O4 | 1.74(5) |
| O5 | 1.78(7) |
| O6 | 1.79(6) |
| O7 | 1.59(6) |

**Table 4:** Basis vectors for the space group $Pbn2_1$ with $k=(0,0,0)$. The decomposition of the magnetic representation for the Co site is $\Gamma_{mag}=3\Gamma_1^1+3\Gamma_2^1+3\Gamma_3^1+3\Gamma_4^1$. The atoms are defined according to 1:$(x,y,z)$, 2:$(-x,-y,z+1/2)$, 3:$(-x+1/2,y+1/2)$ and 4:$(x+1/2,-y+1/2,z+1/2)$.

|  | $\Gamma_1$ | | | $\Gamma_2$ | | | $\Gamma_3$ | | | $\Gamma_4$ | | |
|---|---|---|---|---|---|---|---|---|---|---|---|---|
|  | $\Psi_1$ | $\Psi_2$ | $\Psi_3$ | $\Psi_4$ | $\Psi_5$ | $\Psi_6$ | $\Psi_7$ | $\Psi_8$ | $\Psi_9$ | $\Psi_{10}$ | $\Psi_{11}$ | $\Psi_{12}$ |
| Atom | $m_{//a}$ | $m_{//b}$ | $m_{//c}$ | $m_{//a}$ | $m_{//b}$ | $m_{//c}$ | $m_{//a}$ | $m_{//b}$ | $m_{//c}$ | $m_{//a}$ | $m_{//b}$ | $m_{//c}$ |
| $m_1$ | 1 | 1 | 1 | 1 | 1 | 1 | 1 | 1 | 1 | 1 | 1 | 1 |
| $m_2$ | -1 | -1 | 1 | -1 | -1 | 1 | 1 | 1 | -1 | 1 | 1 | -1 |
| $m_3$ | 1 | -1 | -1 | -1 | 1 | 1 | 1 | -1 | -1 | -1 | 1 | 1 |
| $m_4$ | -1 | 1 | -1 | 1 | -1 | 1 | 1 | -1 | 1 | -1 | 1 | -1 |

**Figure Captions**

**Fig. 1** (Color online) Rietveld refinement of $CaBaCo_4O_7$ at 4 K: The dots and solid line represent the experimental data points and calculated diffraction pattern, respectively. The difference is shown at the bottom. The row of markers shows the positions of the nuclear and magnetic reflections for $CaBaCo_4O_7$ and CoO. The thick solid line (blue online) represents the contribution from magnetic scattering of $CaBaCo_4O_7$ alone. The inset shows the specific heat C versus T.

**Fig. 2** (Color online) Evolution of the cell parameters (a) and of the cell volume and structural distortion ($D = (b\sqrt{3} - a)/a$)(b) versus temperature of $CaBaCo_4O_7$

**Fig. 3** (Color online) Structure of $CaBaCo_4O_7$ at 4 K showing the arrangement of the kagomé (K) and triangular (T) layers: (a) projection along $\vec{a}$, (b) projection along $\vec{b}$.

**Fig. 4** (Color online) Magnetic structure of $CaBaCo_4O_7$: projection along $\vec{c}$ (a) and projection along $\vec{a}$ (b). The cobalt network, built up of "Co5" triangular bipyramids and "Co3" triangles is drawn as a support

**Fig. 5** Evolution of the spontaneous magnetization versus temperature for $CaBaCo_4O_7$. The resulting moment is given in Bohr magneton per f.u. The line is a guide to the eye.

**Fig. 6** (Color online) Evolution of the magnetic moments of the different cobalt ions versus temperature for $CaBaCo_4O_7$. The lines are guide to the eye.

**Fig. 7** Comparison of the magnetic structures of $YBaCo_4O_7$ at 80 K (a) and of $CaBaCo_4O_7$ at 4K (b) view along $\vec{c}$. The ellipses show the similar triple ferrimagnetic chains "Co1 Co2 Co3" and their relative orientations in the two compounds.

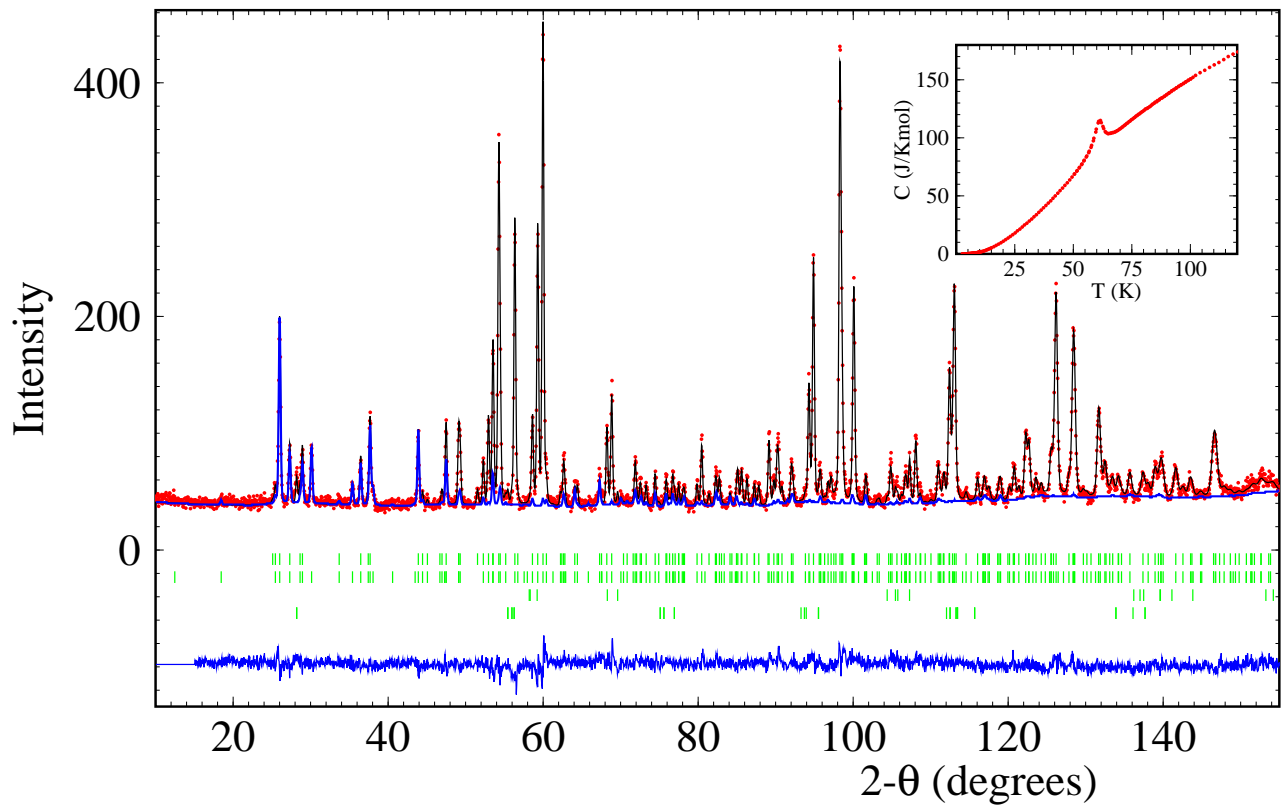

Figure 1

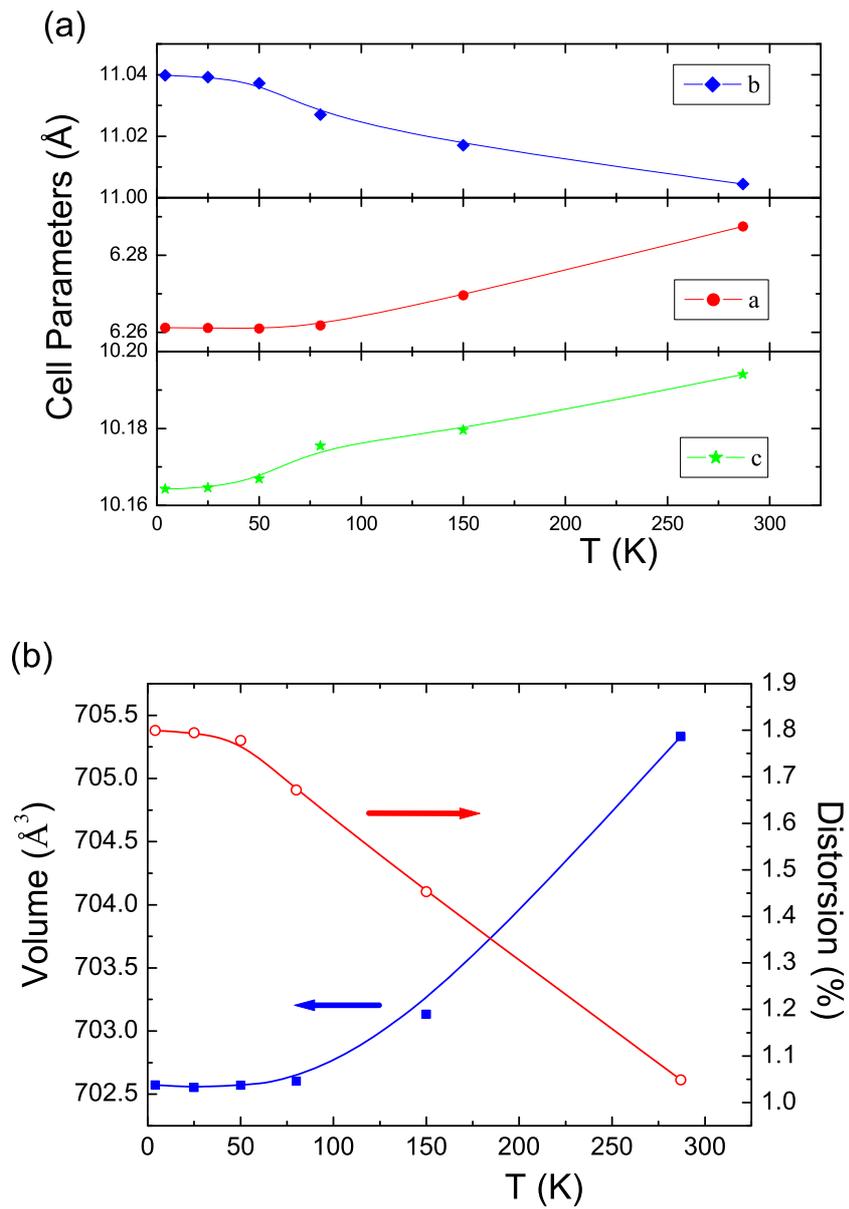

Figure 2

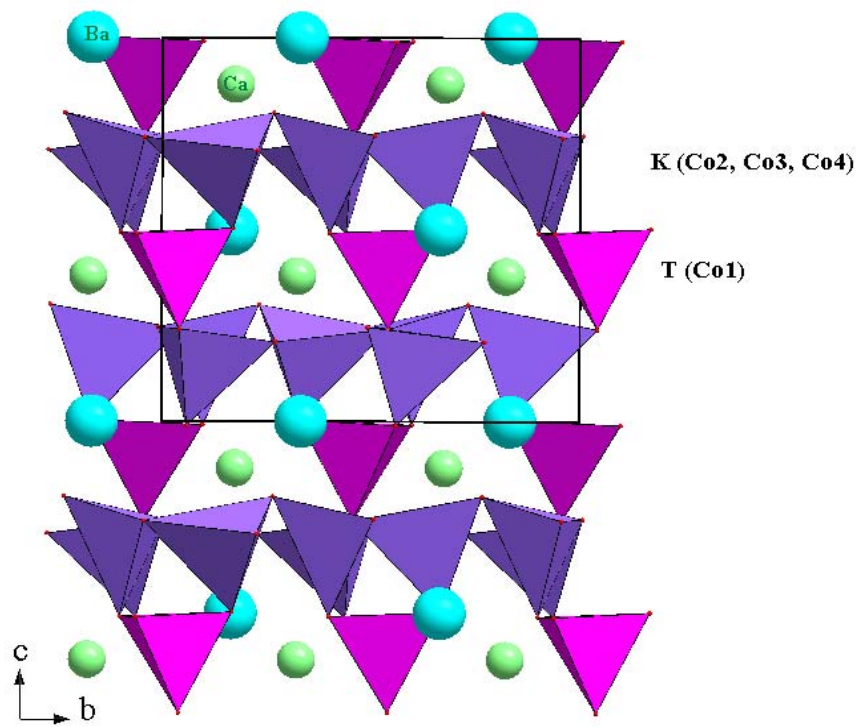

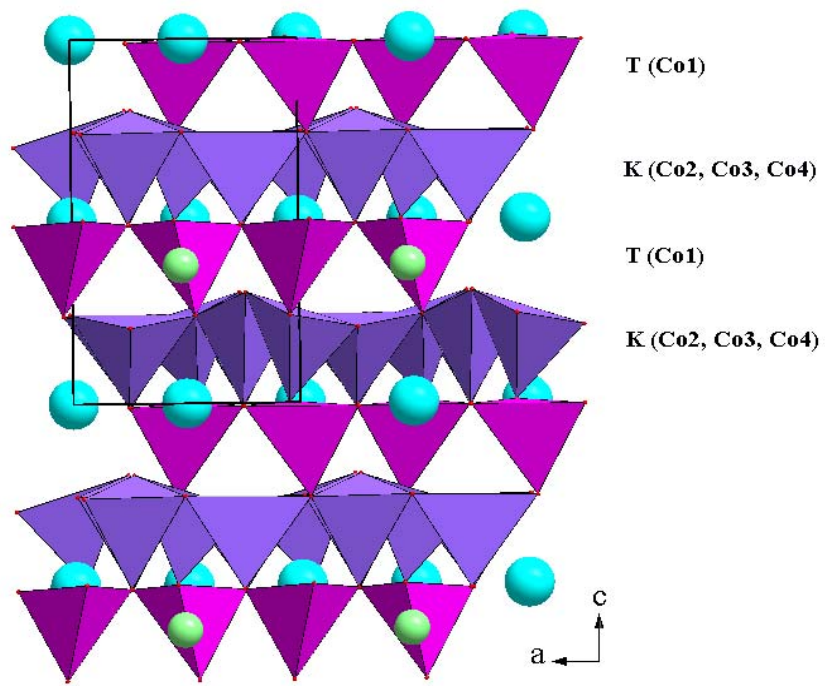

Figure 3

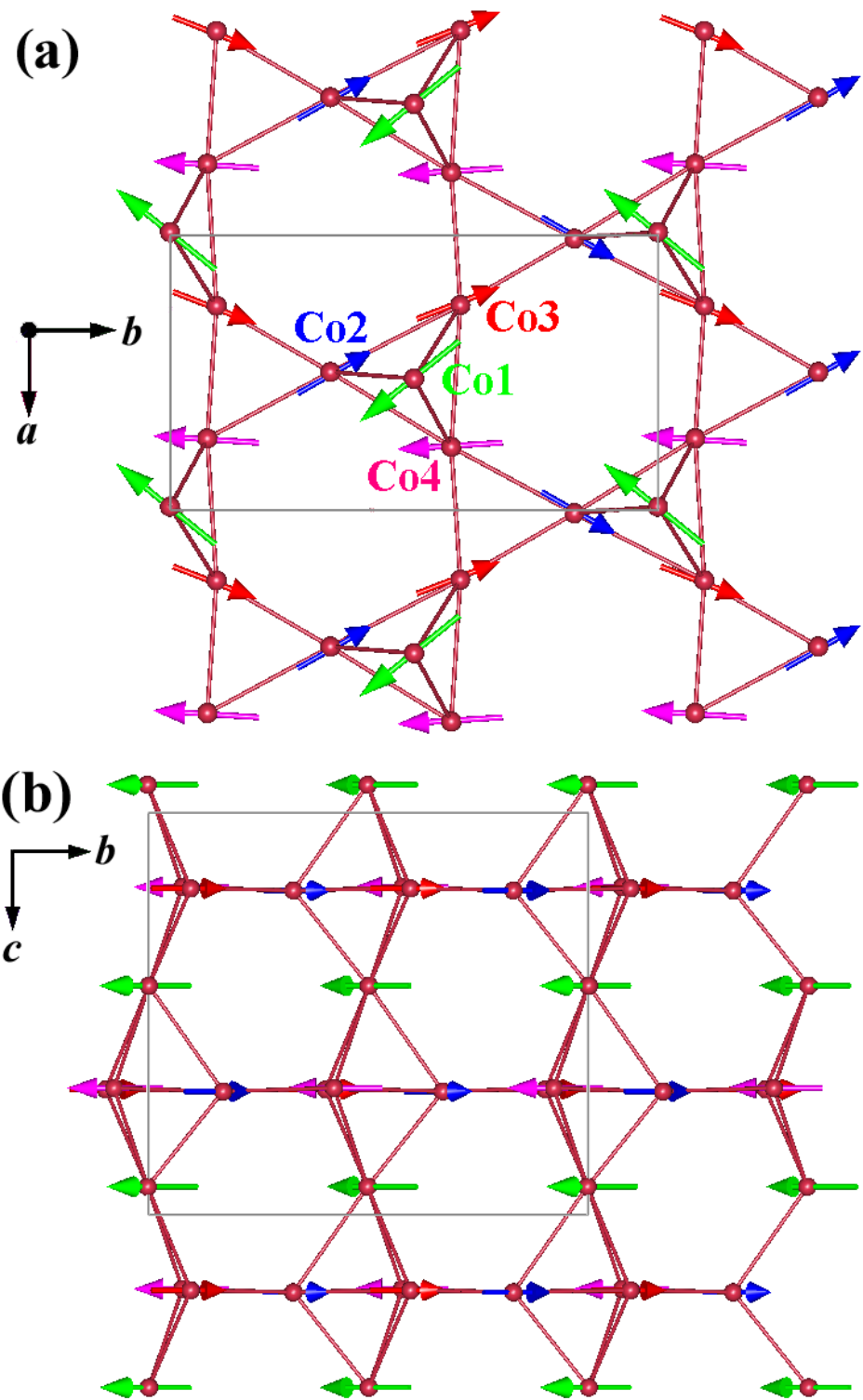

Figure 4

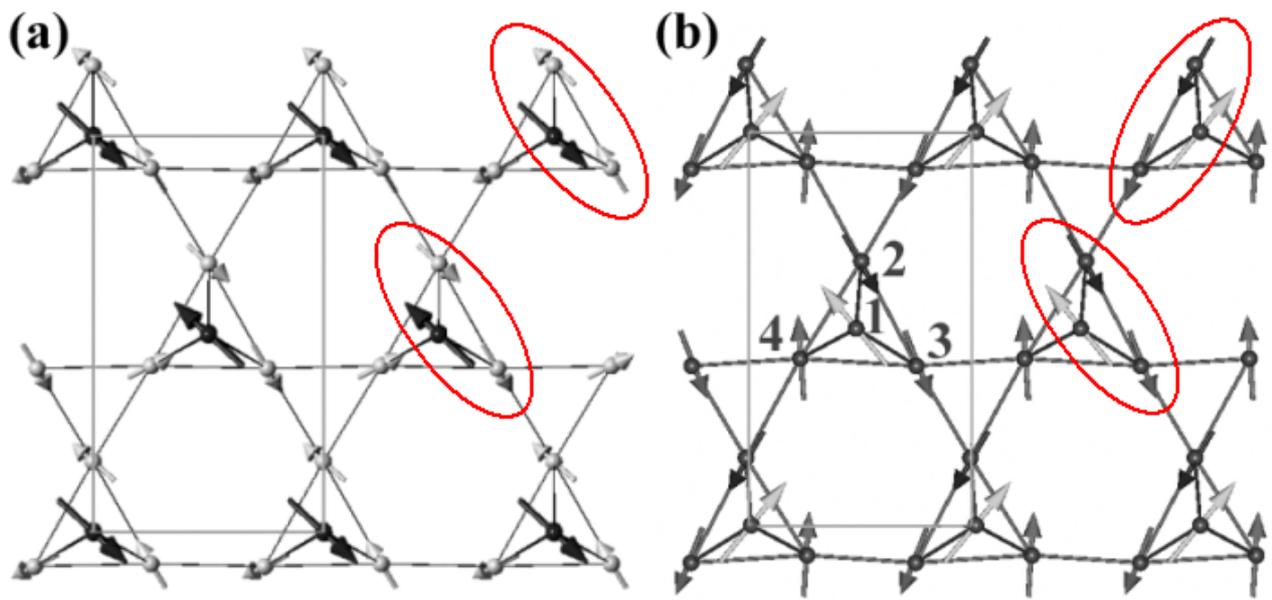

Figure 5

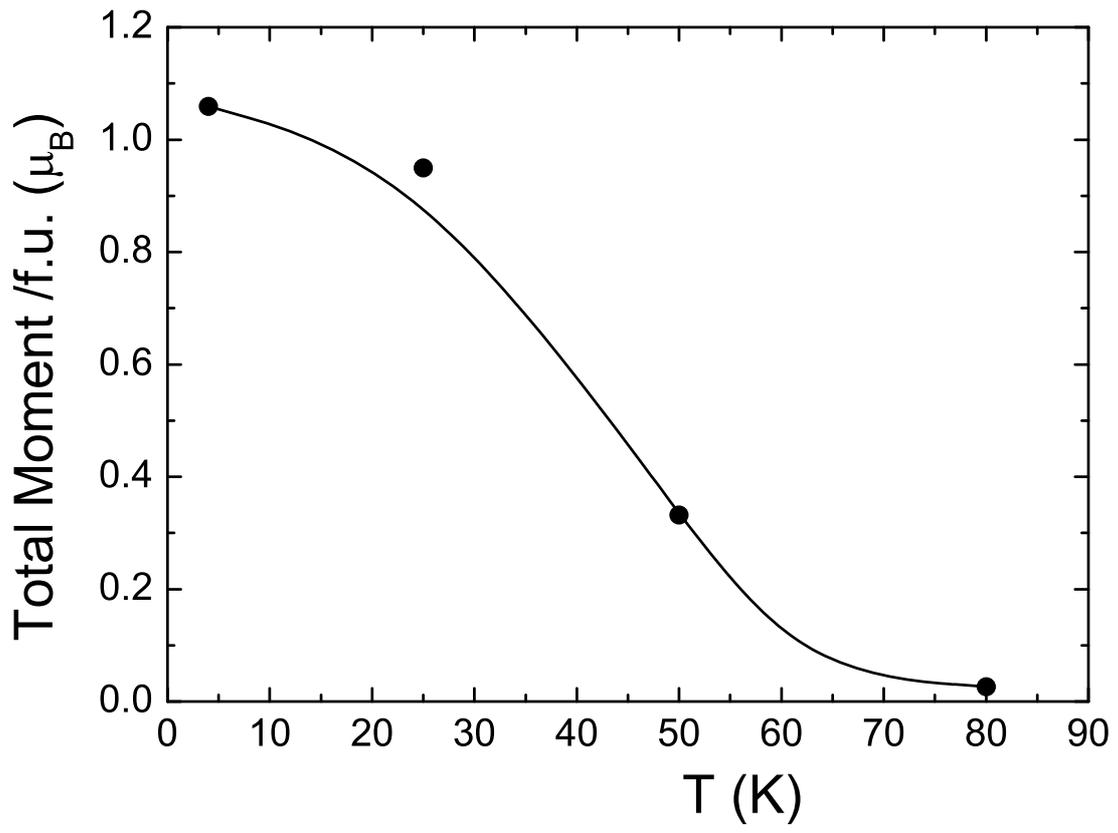

Figure 6

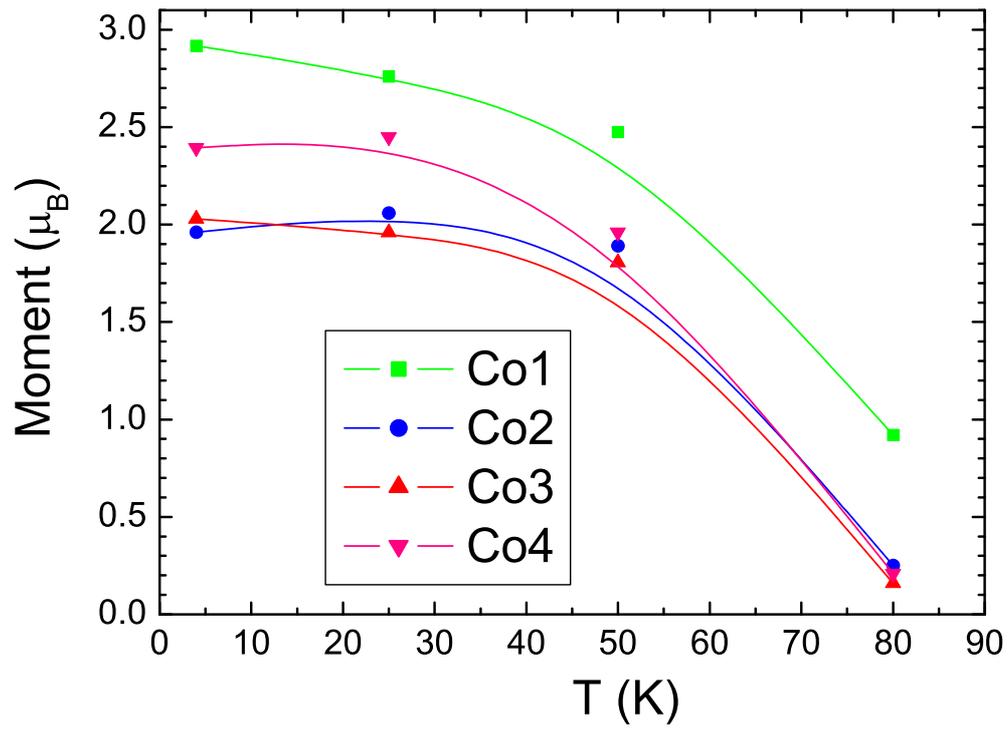

Figure 7